# D³-Guard: Acoustic-based Drowsy Driving Detection Using Smartphones

Yadong Xie∗   Fan Li∗   Yue Wu∗   Song Yang∗   Yu Wang†
∗ School of Computer Science, Beijing Institute of Technology, Beijing, 100081, China.
† Department of Computer Science, University of North Carolina at Charlotte, Charlotte, NC 28223, USA.
∗ {ydxie, fli, ywu, S.Yang}@bit.edu.cn, † yu.wang@uncc.edu

*Abstract*—Since the number of cars has grown rapidly in recent years, driving safety draws more and more public attention. Drowsy driving is one of the biggest threatens to driving safety. Therefore, a simple but robust system that can detect drowsy driving with commercial off-the-shelf devices (such as smartphones) is very necessary. With this motivation, we explore the feasibility of purely using acoustic sensors embedded in smartphones to detect drowsy driving. We first study characteristics of drowsy driving, and find some unique patterns of Doppler shift caused by three typical drowsy behaviors, i.e. nodding, yawning and operating steering wheel. We then validate our important findings through empirical analysis of the driving data collected from real driving environments. We further propose a real-time <u>D</u>rowsy <u>D</u>riving <u>D</u>etection system (D³-Guard) based on audio devices embedded in smartphones. In order to improve the performance of our system, we adopt an effective feature extraction method based on undersampling technique and FFT, and carefully design a high-accuracy detector based on LSTM networks for the early detection of drowsy driving. Through extensive experiments with 5 volunteer drivers in real driving environments, our system can distinguish drowsy driving actions with an average total accuracy of 93.31% in real-time. Over 80% drowsy driving actions can be detected within first 70% of action duration.

## I. INTRODUCTION

Drowsy driving [1] is a significant factor in causing severe traffic accidents. The National Highway Traffic Safety Administration reported that 803 fatalities were drowsy-driving-related in 2016 based on the records in Fatal Accident Reporting System. These reported fatalities have remained largely consistent across the past decade. Between 2011 and 2015, there were a total of 4,121 crashes related to drowsy driving [2]. However, many drowsy driving events are not obvious or recorded so that most drivers fail to realize the risk of drowsy driving. Recent research [3] shows that the attentions on drowsy driving are much less than several other issues, such as using cell phones, speeding, distracted driving and drunk driving. Consequently, it is necessary to develop a drowsy driving detection system which can remind drivers in the early stage to have a rest and thus reduce potential traffic risk.

Although there have been many works [4], [5] on drowsy driving detection, most of them use additional deployed devices to detect driver movements and vehicle status. These devices include infrared sensors, high-definition cameras, and electroencephalograph devices which are expensive to design and deploy. Nowadays, smartphones become powerful with enriched inertial sensors, such as cameras, accelerators and gyroscopes, which can be used to sense various aspects of our lives [6]. Under recent trend of smartphone-based sensing, there emerge a great number of smartphone applications aiming at detecting driving behaviors [7]–[9]. But most of them do not focus on detecting drowsy driving. Therefore, it is an urgent need to develop smartphone-based drowsy driving detection system which does not rely on any extra equipments.

From the aforementioned motivations, we first explore notable characteristics of drowsy driving. According to the explanation of American Academy of Sleep Medicine [10], typical drowsy driving behaviors include inability to keep eyes open, nodding, yawning, missing road signs or turns and drifting into other lanes. Some of these behaviors have been considered in previous drowsy driving detection systems. Sober-Drive [5] based on blinking detection has a good performance on detecting drowsy driving, but it is hard to deploy on all smartphones due to its high demand on hardwares. Kithil *et al.* [11] also find that some people can be functionally asleep with their eyes open and this suggests that detecting the status of eyes is not accurate enough. In contrast, head motion detection and yawning detection are more robust. McDonald *et al.* [12] show that when a driver is fatigued, the time between two continues minor corrections of the Steering Wheel (SW) becomes longer. Meanwhile, large and rapid corrections become more frequent. Based on these and other investigations, we choose not only nodding, but also yawning and operating steering wheel as the features to design our drowsy driving detection system. The goal of our system is to recognize drowsy driving through detecting these three actions in real-time using solely smartphone sensors.

Taking recent advances in acoustic sensing [13]–[15], we only use the audio devices (the embedded microphone and speaker) within smartphones to perform the drowsy driving detection. This makes such solution easy and cheap to be adopted, since every smartphone (or even every phone) has such devices. However, this also poses challenges on the detection system. One of the key challenges is how to effectively identify nodding, yawning and operating steering wheel in various driving motions purely based on obtained acoustic data. After analyzing real driving data, we find that for a particular drowsy driving action, different drivers have similar patterns. According to this finding, we propose a method for detecting drowsy driving actions through collecting Doppler profiles of audio signals when driving. It is well-known that actions of the human body can produce Doppler shift [13], [15]. Thus we first collect audio signals of drowsy driving actions by smartphones. Then we use Fast Fourier Transform (FFT) to extract effective features from Doppler profiles of audio signals. Furthermore, in order to improve the frequency resolution, we adopt undersampling [16] that translates high-frequency bandpass signals into low-frequency lowpass signals without frequency spectrum distortion.

When we get the extracted features, another challenge arises: how to produce accurate identification results quickly? To solve this, we employ a deep learning method, called Long Short Term Memory (LSTM) networks, to generate detectors for each drowsy driving action. LSTM is a kind of Recurrent Neural Networks (RNNs) with principally infinite memory for every computing node. Similar to other RNNs, LSTM is predestined for analyzing sequential data like audio signals, but it also has inherent advantages in early detection. To meet realistic demands, we collect training set data using smartphone sensors for 6-month of five drivers in real driving environments. In addition, we also supply with some simulated data by these drivers to enrich the training set. Experiment results show that our system is reliable and efficient in real driving environments.

Our contributions are summarized as follows:

- To the best of our knowledge, we are the first to study the unique patterns of Doppler shift caused by drowsy driving. We find that for a particular drowsy driving action, different drivers have similar patterns. We verify this finding by analyzing the driving data collected in real driving environments.
- We propose a new method, D$^3$-Guard, for detecting drowsy driving actions based on audio devices embedded in smartphones. It can detect nodding, yawning and abnormal operating steering wheel in real time. Furthermore, D$^3$-Guard adopts a high-accuracy detector based on LSTM networks to early detect drowsy driving.
- We conduct extensive experiments both in real and simulated driving environments. The results show that D$^3$-Guard can distinguish drowsy driving actions in real-time. And it achieves an average total accuracy of 93.31% for drowsy actions detection, and over 80% drowsy driving actions can be detected within the first 70% of action duration.

The remainder of the paper is organized as follows. We review related work in Section II. In Section III, patterns of drowsy driving actions on Doppler profiles are analyzed. Section IV presents the detailed design of D$^3$-Guard. Implementation and extensive experimental results are provided in Section V. Finally, we draw our conclusion in Section VI.

## II. RELATED WORK

In this section, we present current work relevant to drowsy driving detection. Specifically, we review *specific detection device* and *smartphone* based approaches.

**Specific detection device based solution.** Electroencephalography (EEG) headset is used to collect EEG signals of drivers which can evaluate the drowsiness stage [4]. EyeAwake [17] leverages infrared sensors consisting of an infrared Light Emitting Diode (LED) and an infrared phototransistor to measure the eye blinking rates of drivers. A camera-based real-time driver-fatigue monitor [18] detects drowsy driving by extract visual cues. A system of detecting lane departures related to drowsiness [12] uses steering wheel angle data collected by a simulator. However, these works are all based on additional deployed devices (many of them are customized) that need high cost and are easily affected by the environment.

**Vision-based smartphone approaches.** Many researches leverage smartphone cameras for drowsy driving detection. A driver attention detection system [19] focuses on estimating driver gaze direction through smartphone videos. Sober-Drive [5] analyzes blink time and blink rate to detect drowsy driving by using front camera of smartphones. A drowsy monitoring system [20] captures facial expressions like eye blinking, head shaking and yawning to judge the vigilance level of drivers. CarSafe [21] monitors and detects whether the driver is tired or distracted using the front camera while tracking road conditions using the rear camera on smartphones at the same time. However, the accuracy of these camera-based approaches depending on visibility is often unreliable, since these approaches depend on conditions like lighting and weather. And there needs at least one camera facing the driver.

**Other smartphone based approaches.** There are also some researches focusing on driving behavior detection using other smartphone inertial sensors. SenSpeed [7] estimates a vehicle speed through sensing natural driving conditions in urban environments with accelerators and gyroscopes of smartphones. V-Sense [22] detects various vehicle maneuvers, including lane-changes, turns, and driving on curvy roads, by only utilizing non-vision sensors on the smartphones. TEXIVE [8], [9] uses smartphone sensors to distinguish drivers from passengers, and detects texting operations during driving. ER [23] detects inattentive driving events, such as turning back and picking up drops, at an early stage leveraging audio devices. Our work is different from the previous studies in that the amplitudes of drowsy driving actions are much smaller than those of other driving actions. And the time duration of each drowsy driving action is relevantly short. Thus, we need more powerful abilities of sensing and detection for them. To the best of our knowledge, there is no acoustic-based drowsy driving detection using the fusion of multiple action features on smartphones.

## III. PRELIMINARY

In this section, we first introduce the patterns of our three selected drowsy driving actions, and then analyze the features of Doppler shifts caused by these actions.

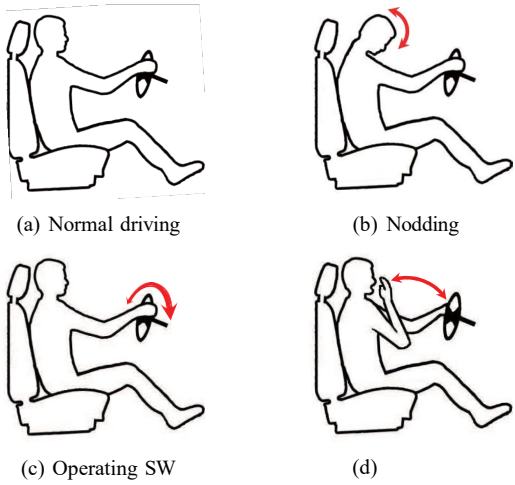

Fig. 1: Normal and drowsy driving actions.

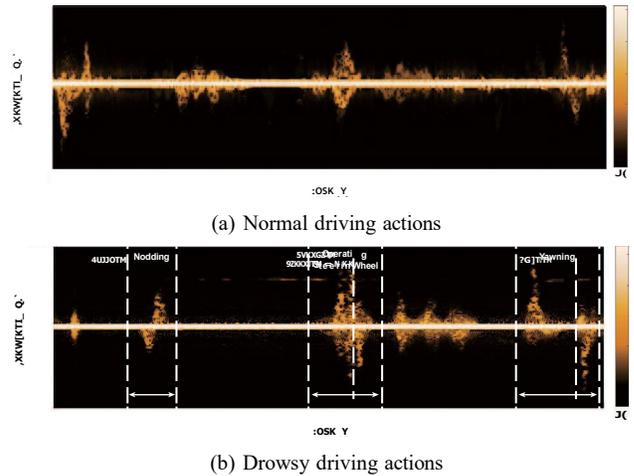

Fig. 2: Frequency-time Doppler profiles of normal and drowsy driving actions.

*A. Patterns of Drowsy Driving Actions*

Drowsy drivers put themselves and other pedestrians in a dangerous situation. There are three types of common drowsy driving actions of drivers, as shown in Fig. 1.

**Nodding:** When a driver drives normally, his/her head always has a small gap with the headrest, since it is inconvenient for a driver to observe road conditions that are close to the vehicle if resting his/her head on the headrest, as shown in Fig. 1(a). But when the driver is fatigue, he/she usually has nodding actions, appearing as a quick bow and then looking up, as shown in Fig. 1(b). Through our observations in real road testing, if a driver rests his/her head on the headrest first, resting head eventually changes into nodding.

**Operating Steering Wheel (SW):** When a driver drives normally, he/she usually adjusts the steering wheel smoothly and slowly even on straight roads. But with the deepening of drowsiness, the ability to control vehicles gradually decreases, namely, the frequency of adjusting steering wheel decreases. Then the driving direction cannot be corrected in time, resulting in vehicles moving laterally for a longer distance. When the driver suddenly realizes this situation, he/she usually makes a quick and large adjustment of the steering wheel to correct the driving direction, as shown in Fig. 1(c).

**Yawning:** Yawning is a kind of frequent action when people get fatigue. Usually yawning includes two types of actions. One type is opening mouth wide and then closing it. The other is covering mouth by a hand when opening mouth and then putting down the hand. The second type is shown in Fig. 1(d).

Through analyzing the above three drowsy driving actions, we realize that each action is not a transient but consecutive action lasting for a period of time. Our work is to detect these consecutive drowsy driving actions in real time and try to alert drivers as early as possible.

*B. Action Patterns via Doppler Shifts*

We use Doppler shift of acoustic signals to capture the patterns of driver actions. Doppler shift refers to the change in frequency or wavelength of audio signals in relation to observer who moves relatively to the signal source. The frequency offset $\Delta f$ is determined by the relative velocity $\Delta v$ between the source and the observer. Formally, $\Delta f = (\Delta v/c) \cdot f_0$, where $f_0$ is the emitted frequency and $c$ is the speed of waves.

We can find that a higher frequency $f_0$ results in a more obvious Doppler shift. According to [16], the sound frequency above $15kHz$ is already inaudible for most adults, but some minority groups like young kids who are more sensitive to high frequency sounds can even hear sound up to $18kHz$. Since most commodity smartphone speakers can only produce sound frequency at up to $20kHz$, we select audio signal with fixed frequency $f_0 = 20kHz$. In addition, we set $f_s = 44.1kHz$ (supported by most mainstream smartphones), which is the default sampling rate of audio signals under $20kHz$. Then, to extract effective features from Doppler profiles of audio signals, we use 2,048-points FFT to get frequency domain information. Fig. 2 shows the Doppler profile structures of normal driving and drowsy driving on the same road. We can see that there are some similarities between these two structures, since they have both positive and negative Doppler shifts. However, the differences are more noticeable such as frequency range and energy amplitude. In Fig. 2(b), we can also find that the Doppler profile pattern of nodding differs from yawning and operating steering wheel obviously. Thus we can detect and distinguish these three actions to detect drowsy driving.

From the analysis over more real road driving data, we find that each drowsy driving actions has unique patterns on the structure of Doppler profiles. This suggests that Doppler shift of audio signals caused by drowsy driving actions has great potential to be used in drowsy driving detection.

## IV. SYSTEM DESIGN OF $D^3$-GUARD

We now show the design of $D^3$-Guard, our drowsy driving detection system, which leverages Doppler shift of audio signals to capture the unique patterns of drowsy driving actions.

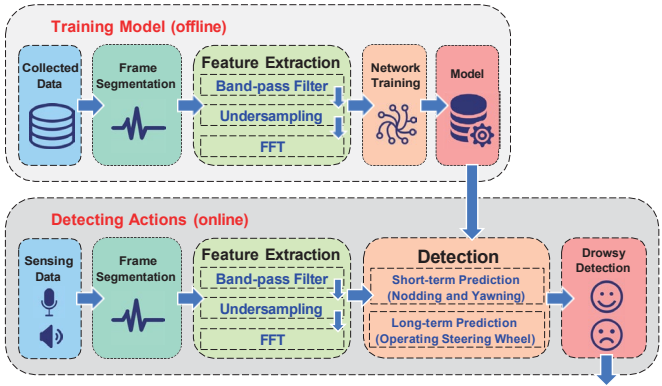

Fig. 3: System architecture and work flows in D³-Guard.

*A. Overview*

Fig. 3 shows the architecture of D³-Guard. The whole system includes two parts, *training model for drowsy driving actions* (offline phase) and *detecting drowsy driving actions* (online phase).

In the offline phase, we first collect labeled driving data in real driving environments and split the audio signals into suitable frames. Then, we extract effective features from the Doppler profiles of audio signals for each drowsy driving action. After this, we train these features through deep learning methods to generate two classifiers for short-term prediction and long-term prediction. We propose two LSTM networks, one network is responsible for a short-term prediction, including nodding and yawning. The other network is responsible for a long-term prediction, including operating steering wheel. Afterwards, the two prediction results are turned into a drowsy index through a simple Deep Neural Network. Finally, we store these networks in the database for online phase (i.e., detecting drowsy driving actions in real time).

In the online phase, D³-Guard uses microphones to collect audio signals generated by speakers in real time, and splits the audio signals into suitable frames. Then, to extract effective features, the audio signals are transformed to Doppler profiles through FFT. We also adopt band-pass filter and undersampling technique (both in offline and online phases) to improve frequency resolution without distorting frequency spectra. After that, we send the processed frames to LSTM networks, and each LSTM network considers not only the current frame but also several previous frames to give a prediction. The two prediction results are turned into an index through the Deep Neural Network. Finally, D³-Guard determines whether a driver is fatigue or not according to the index. Once D³-Guard identifies drowsy driving through the above procedures, it sends a warning message to alert the driver.

*B. Modeling Drowsy Driving Actions*

**Collecting Data and Segmenting Frames:** The training of networks requires a large number of training data. To collect training data, we develop an Android application to generate and collect audio signals. We recruit five drivers (3 males and 2 females) to collect data by driving different vehicles. The drivers carry different smartphones, such as SAMSUNG GALAXY Note3, HUAWEI nova2 Plus and HTC One M9. In order to obtain ground truth, we equip all vehicles with cameras to capture the states of these drivers. And the drivers are asked to record whether they are fatigue after each driving. The 5 drivers collect two types of training data, including real driving data and simulated data. We collect real driving data from September 13, 2017 to March 21, 2018. The driving data keeps being collected whenever the drivers are commuting or driving long-distance. The drivers have different driving habits and driving routes. As a supplement for real driving data, we also collect simulated data. To be specific, volunteer drivers imitate the three drowsy driving actions in normal driving conditions. After collecting data, we distinguish drowsy driving actions manually from the 6-month data through checking videos recorded by the cameras as ground truth. Finally, we obtain 5,812 samples of drowsy driving actions from the real driving data and 8,336 samples from the simulated data. Finally, we use the processed audio signals with labels to construct a training dataset $X = \{N, S, Y\}$, which contains 3 types of drowsy driving actions. $N$, $S$, $Y$ indicate the sample sets of nodding, operating steering wheel samples and yawning, respectively.

We cut a sample directly into several segments with fixed length, each segment is a frame. We also divide each action into several steps. We label each frame with the action and the step it belongs to. These frames are used later for extracting features and training model.

**Extracting Features:** According to the analysis in Section III, each drowsy driving action has its unique pattern in frequency and time domain. The main information stored in raw signals is temporal information, but it is difficult to get effective features of drowsy driving actions in the time domain. In contrast, we can get effective features in the frequency domain, so it is more efficient to take frequency-domain information as the input of training networks.

In our feature extraction algorithm, we first use a band-pass filter. Through our observations, we find that normal drowsy driving actions lead to a Doppler shift ranging from $-200Hz$ to $200Hz$. Therefore, we adopt the filter for audio signals to obtain the target frequency band ranging from $19.8kHz$ to $20.2kHz$. After filtering, we can eliminate other out-band interferences and prepare for undersampling.

D³-Guard detects drowsy driving through analyzing Doppler frequency shifts caused by body actions. However, a drowsy driving action usually does not last very long (i.e. generally shorter than $3.5s$), and the amplitude is not very large, which lead to limited Doppler frequency shifts. Sun *et al.* [24] indicate that there are three options to improve the situation: 1) increasing audio frequency $f_0$, 2) adding FFT points, and 3) decreasing sampling rate $f_s$. The first two options both have drawbacks. For the first option, most commodity smartphone speakers can only produce sound frequency at up to $20kHz$. For the second option, adding FFT points would bring a higher computational burden, so we adopt the last option. Nyquist sampling theorem states that the sampling

1228

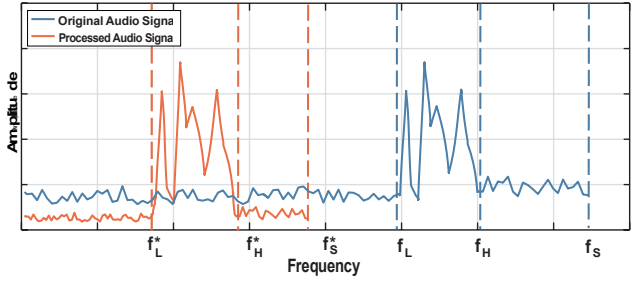

Fig. 4: The effect of undersampling to the spectrums.

TABLE I: Possible $f_S^*$ and range of $f_L$ under different $n$.

| $f_S(kHz)$ | $n$ | $f_S^*(kHz)$ | Range of $f_L(kHz)$ |
|---|---|---|---|
| 44.1 | 4 | 11.0 | (16.5, 20) |
| 44.1 | 5 | 8.8 | (17.7, 20) |
| 44.1 | 6 | 7.3 | (18.3, 20) |
| 44.1 | 7 | 6.3 | (18.9, 20) |
| 44.1 | 8 | 5.5 | (19.3, 20) |

frequency has to be twice the maximum signal frequency, otherwise the sampled signal would be aliased. However, if the bandwidth of a bandpass signal is significantly smaller than the central frequency of the signal, it is still possible to sample the signal at a much lower rate than the Nyquist sampling rate, without causing frequency alias. We can address this challenge by utilizing undersampling technique. The effect of undersampling technique on spectrums is shown in Fig. 4. It can be seen that the original audio signal is shifted from high frequency to low frequency after being processed by undersampling.

We use $f_L$ and $f_H$ to denote the lowest and the highest band limits of the received frequency-shifted signal, then the bandwidth of the signal $B = f_H - f_L$. According to undersampling theorem, shifts of bands from $f_L$ to $f_H$ must not overlap when shifted by all integer multiples of new sampling rate $f_S^*$. This can be interpreted as the following constraint:

$$\frac{2 \cdot f_H}{n} \leq f_S^* \leq \frac{2 \cdot f_L}{n-1}, \forall n : 1 \leq n \leq \lfloor \frac{f_H}{B} \rfloor, \quad (1)$$

where $\lfloor \cdot \rfloor$ is the flooring operation and $n = f_S/f_S^*$ is the undersampling factor. Table I summarizes the possible $f_S^*$ supported in our system under different undersampling factors $n$. We can see that $f_S^*$ decreases when $n$ increases, which leads to better feature extraction results. Therefore, we apply $n = 8$ for the optimal setting of frequency resolution, corresponding to an undersampling sampling rate $f_S^* = 5.5kHz$.

After the undersampling process, the spectra of original audio samples are shifted from $19.80kHz - 20.20kHz$ to a much lower spectrum of $1.80kHz - 2.20kHz$ with a central frequency at $2kHz$. Finally, to obtain frequency shifts, each undersampled audio frame is processed by a 2,048-point FFT, which achieves a high frequency resolution with an appropriate computational complexity. The results of FFT are complexes which contain a real part and an imaginary part. We can get the amplitude and phase information from the real and imaginary parts. Through a large number of experiments, we find that using the phase information to train networks has a better effect than the amplitude. Therefore, we take the phase information as our feature vectors which are transformed from audio signal frames.

**Training Model:** After getting feature vectors, D3-Guard uses LSTM networks [25] to train the classifier model for each drowsy driving action. Based on the training results, we construct a DNN to produce the final results.

Traditional modeling methods usually treat separate frames of sensing data as statistically independent. Feature vectors converted from independent frames usually lack temporal context. However, ignoring the temporal context during modeling may limit the performance of our system in some challenging tasks. In order to exploit the temporal dependencies within our sensing data, we adopt LSTM networks. This architecture is recurrent, where LSTM networks consider not only the current frame but also several previous frames. Each LSTM cell keeps tracking an internal state that represents *memory*. With this, LSTM networks have abilities to retain information across tens of frames. LSTM networks have certain timesteps and each timestep takes a frame as input. Then the LSTM networks produce a classified result at each timestep.

According to the analysis of training dataset $X$, we find 95% of nodding, yawning and operating steering wheel can complete in $2.3s$, $2.7s$ and $2.4s$. We use $T(N)$, $T(Y)$ and $T(S)$ to denote the time duration from the beginning to the end of nodding, yawning and operating steering wheel. For convenience we call the time duration as total time. Through the above analysis, we set $T(N) = 2.3s$, $T(Y) = 2.7s$ and $T(S) = 2.4s$. The total time of three drowsy driving actions are used to estimate the timeliness of our system in Section V. We divide samples into frames before extracting features, and we find that a frame equaling to $0.25s$ can achieve a good performance through experiments.

We design two LSTM networks in D3-Guard with different structures. One network is responsible for a short-term prediction called LSTM-S. It has 11 timesteps which means it focuses on the latest $2.75s$ of the driver's state. It has two LSTM layers and one softmax layer. Each LSTM layer can convert the input features to a set of compressed representations through an unsupervised manner. Such compressed representations are able to characterize unique driving actions. At $t$-th timestep, the LSTM layers can map the input $x_t$ of $n$-th sample into a compressed vector $h_t$ as below:

$$h_t = \sigma(W_o [h_{t-1}, x_t] + b_o) \cdot tanh(C_t), \quad (2)$$

where $W_o$ and $b_o$ denote a weight matrix and a bias vector of output gate, and $C_t$ represents the status at $t$-th timestep. Given $h_t$, if we have $K$ types of actions, the class probability vector $P_t = [P_t^1, P_t^2, \cdots P_t^K]$ is calculated as

$$P_t = s(W^T h_t + b), \quad (3)$$

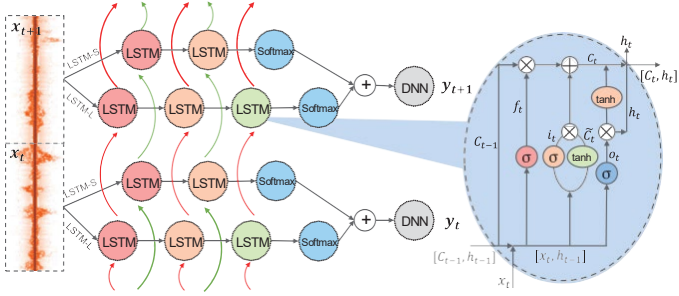

Fig. 5: Overall network architecture and the internal structure of one LSTM unit.

where $s(\cdot)$ is a softmax function, $W^T$ is a weight matrix and $b$ is a bias vector. The class label $l_t$ at $t$-th timestep is then assigned to the class with the highest probability

$$l_t = \arg\max_j P_t^j, j \in [1, K]. \tag{4}$$

The network is trained to minimize the differences between ground truth and predictive result. Specifically,

$$\min(c_t, c_t^0) = \min\left(-\sum_{i=1}^{M}(c_t \log c_t^0)\right), \tag{5}$$

where $M$ is the number of training samples, and $c_t^0$ is the output of the network. LSTM-L, the other network, mainly detects abnormal operating steering wheel. Sometimes operating steering wheel by unwearied drivers may also be completed in 2.4$s$ in situations like lane-changing and overtaking. On the basis of [26], in drowsy driving scenes, steering wheel angles usually remain unchanged for a period of time for at least 3$s$ to 4$s$, followed by a quick operating steering wheel with big amplitude. It is necessary to take a little longer time (more than 2.4$s$) to detect abnormal operating steering wheel. Therefore, LSTM-L is responsible for a long time prediction. It has three LSTM layers and one softmax layer. We use the same method to train LSTM-L. The difference between the two networks is that LSTM-S has 11 timesteps while LSTM-L network has 28 timesteps. After training LSTM-S and LSTM-L, we propose a two-layer DNN to produce the final result. LSTM-S and LSTM-L produce one result respectively for each frame. DNN takes these two results as input and the driver's fatigue condition as label. The training method of DNN is similar to that of LSTM network. The DNN can map input $D$ into a possibility of drowsy driving $R = \sigma(W_d D + b_d)$, where $\sigma(\cdot)$ is sigmoid function defined as $\sigma(x) = 1/(1 + e^{-x})$, $W_d$ is a weight matrix and $b_d$ is a bias vector. The objective of the DNN is similar with that of the LSTM network.

Fig. 5 shows the overall architecture of our neural network, which consists of three parts, LSTM-S, LSTM-L and DNN. In LSTM-S, the first layer takes frequency domain information as input and outputs coarse-grained features of short-term actions. The second and third layers take the output of previous layer as input and output fine-grained features. It is notable that each LSTM unit also accepts the state information of previous frame. The detailed working process of producing $h_t$ and $C_t$ from $x_t$, $h_{t-1}$ and $C_{t-1}$ can be seen in zoomed part of Fig.5. Then the softmax layer produces a short-term detection result. Between any two layers, we have a Batch Normalization, which can speed up training, improve model accuracy and prevent overfitting. LSTM-L have similar structure to LSTM-S. But their units have different parameters since they have different timesteps and training datasets. Finally the DNN layer takes the two results of LSTM-S and LSTM-L and produces the probability whether the driver is drowsy driving.

### C. Detecting Drowsy Driving at Online Phase

**Sensing Data and Extracting Features:** These steps are the same with those in offline phase. To detect drowsy driving actions, D³-Guard continuously emits high-frequency sound through smartphone speakers and receives the sound by microphones. The received sound is segmented into frames every 0.25$s$, i.e., each frame contains 0.25-second audio signal. D³-Guard sends the frame into the band-pass filter immediately. Then the filtered frame is processed with undersampling technique to improve its frequency resolution. Next the temporal information of the frame is transformed into frequency domain information through FFT. D³-Guard constructs a feature vector for the frame from its frequency domain information. The above steps keep going continuously when smartphones collect audio signals.

**Online Detection:** The trained networks in the offline phase are stored in smartphones. D³-Guard first sends feature vectors into LSTM networks. LSTM-S is responsible for short-term prediction and it can detect nodding and yawning. It takes the current frame and previous 10 frames as input every 0.25$s$. LSTM-L is responsible for long-term prediction and it can detect abnormal operations of steering wheel. LSTM-L takes the current frame and previous 27 frames as input every 0.25$s$. Findings in [27] suggest the possible enhancement of inference robustness and resource usage level if DNNs are applied to various mobile sensing tasks. Thus we send the two detection results to DNN which produces a probability to indicate whether the driver is fatigue. Once the probability is greater than a specific threshold, D³-Guard sends an alert to the driver.

## V. IMPLEMENTATION AND EVALUATION

In this section, we introduce the implementation details and provide the evaluation results.

### A. Experiment Setup

We implement D³-Guard as an Android application on different smartphones with different versions of Android, and test its performance in real driving environments. To collect valid and enough test data, we invite 5 volunteers (3 males and 2 females) to collect driving data for evaluation. All of them have no deficiency of driving. They drive different vehicles, including electric vehicles and gasoline vehicles. In order to obtain ground truth, all vehicles are equipped with a camera to capture actions of drivers. And the drivers are asked to record whether they are fatigue after each driving. The drivers take tests when they are actually fatigue on empty roads and have a speed limitation of $40km/h$ for safety. Our test scene is shown in Fig. 6. Fig. 6(a) presents an example test area in which there are few pedestrians and vehicles. The drivers can take any routes in this area. Placement locations of smartphones and the camera in vehicles are marked in Fig. 6(b), where smartphones can be placed at four locations. Finally, we distinguish drowsy driving actions manually from the driving data through checking the recorded videos as ground truth.

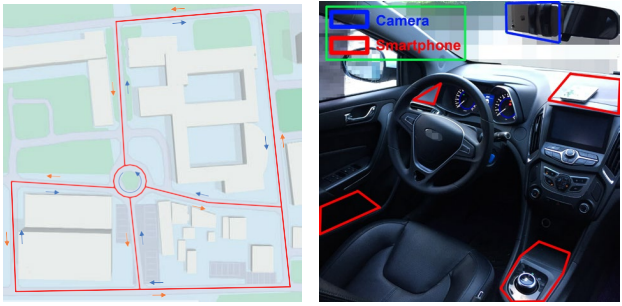

(a) Example of test road skeleton  (b) Smartphone/camera placement

Fig. 6: Test scenarios: (a) an example of test road skeleton, (b) four placement locations of smartphones and one of camera.

### B. Evaluation Methodology

We mainly evaluate $D^3$-Guard from the following aspects.
- *Accuracy:* The percentage of samples which are correctly classified in total samples.
- *Recision:* The percentage of samples which are correctly classified into action A in all samples classified into A.
- *Recall:* The percentage of samples which are classified into action A in the samples truly belong to A.
- *False Alarm:* The percentage of samples which are mistakenly classified into action A in all samples do not belong to A.
- *Missing Alarm:* The percentage of samples which are not classified into action A in the samples truly belong to A.

### C. Overall Performance

We first evaluate the overall performance of $D^3$-Guard. $D^3$-Guard determines whether there is drowsy driving, mainly depending on the detection of three actions. Fig. 7 shows the detection accuracy of different actions and overall drowsy detection for the 5 drivers. $D^3$-Guard achieves an average accuracy of 93.31% for detecting all types of drowsy driving actions. Furthermore, the accuracy of a specific action for all drivers are different, since various drivers have differences in doing the same action. Among the 5 drivers and the 3 actions, the lowest accuracy is 89.57% (Driver 2 and operating SW). Based on the high accuracy of action detection and the integration of 3 action results, we achieve a high accuracy of overall drowsy detection, which is no less than 94.31%.

Fig. 8 shows the precision and recall for different types of drowsy driving actions and drowsy detection. The precision of the 3 types of actions are no less than 92.38%, while the recall of the 3 types of actions are no less than 95.96%. It is easy to see that precision and recall are also very high for drowsy detection.

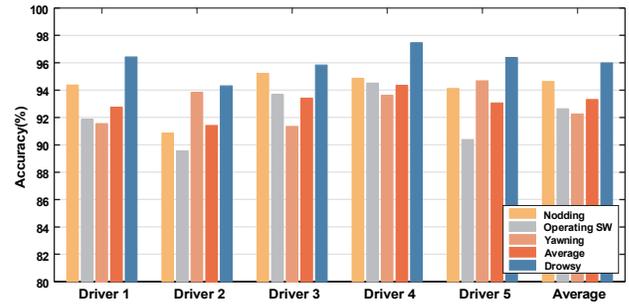

Fig. 7: Accuracy of different actions for 5 drivers.

Fig. 9 shows the false alarm and missing alarm of each type of drowsy driving action and drowsy detection. The false alarm of action detection and drowsy detection are no more than 3.06%, and the missing alarm are no more than 3.52%. These indicate that $D^3$-Guard can reliably alert drivers if he/she is fatigue.

We also analyze the detection time of each correctly detected drowsy driving action, as shown in Fig. 10. It is shown that 87.75% of nodding, 85.38% of yawning and 88.71% of operating steering wheel are detected in $1.75s$, $2s$ and $1.75s$. As the total time of the 3 actions are $2.3s$, $2.7s$ and $2.4s$, $D^3$-Guard can detect at least 80% actions correctly before about 70% of total duration and at least 60% actions correctly before about 50% of total duration. These demonstrate $D^3$-Guard can detect drowsy driving in real time.

### D. Impact of Smartphone Location

Different driver likes to put his/her smartphone in some specific positions in the car. These positions include dashboard (left), dashboard (right), door panel and storage space near gearstick, as shown in Fig. 6(b). Fig. 11 shows the accuracy of different detection targets at four placement locations. It is obvious that $D^3$-Guard has high accuracy at different locations. Most drivers are used to put their smartphones on the left or right of dashboard, this makes the smartphone face to the driver. In these two cases, our system can achieve the highest accuracy. When smartphone is put on the storage space near gearstick, the accuracy of detection is the lowest. This is because this location is far away from the head and has more interference. But at every location, the accuracy of action detection and drowsy detection is no less than 89.43% and 94.84%.

### E. Impact of Network Structure

We design 5 network structures for our experiment:
- *2-LSTM-DNN:* It has 2 LSTM layers and a softmax layer, followed by a 2-layer DNN.
- *3-LSTM-DNN:* It has 3 LSTM layers and a softmax layer, followed by a 2-layer DNN.
- *2-2-LSTM-DNN:* It has 2 LSTM networks with same structures, one has 2 LSTM layers and a softmax layer which detects nodding and yawning, the other detects operating steering wheel, followed by a 2-layer DNN.

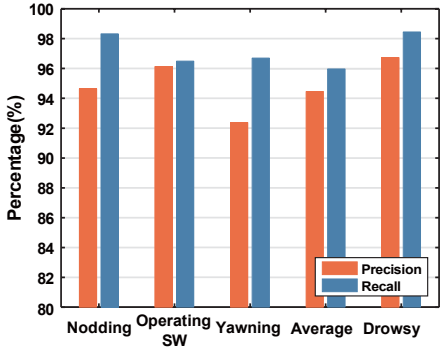

Fig. 8: Precision and recall for different drowsy driving actions.

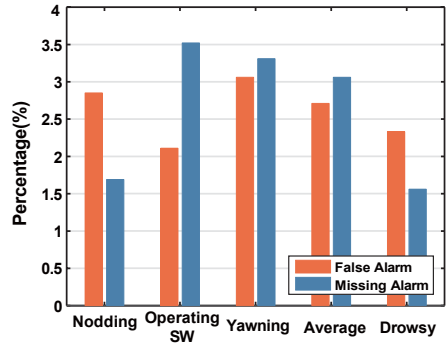

Fig. 9: False alarm and missing alarm for different drowsy driving actions.

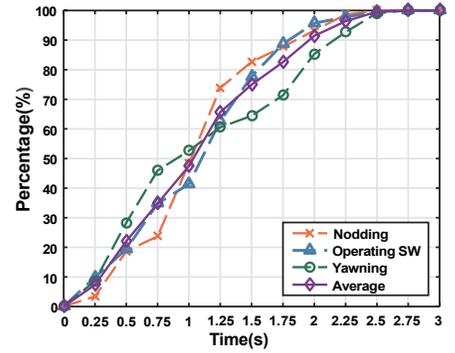

Fig. 10: Detection time for different drowsy driving actions.

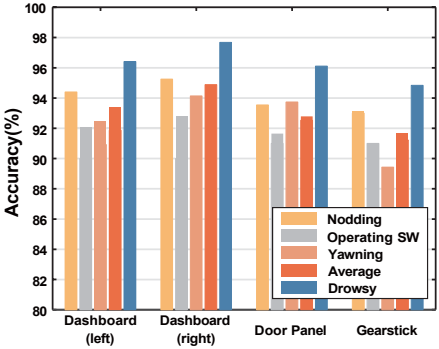

Fig. 11: Accuracy for different locations of the smartphone.

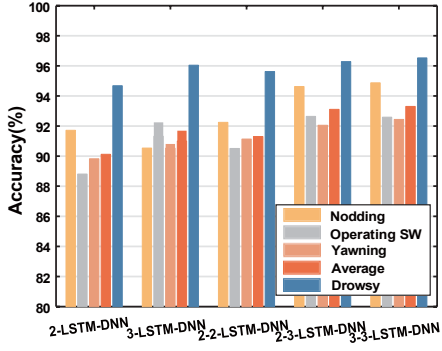

Fig. 12: Accuracy for different network structures.

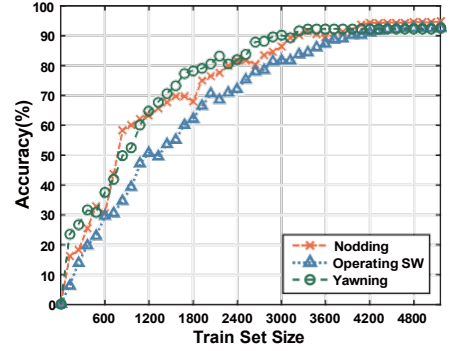

Fig. 13: Accuracy for different training dataset size.

- 2-3-*LSTM-DNN:* It has 2 LSTM networks, one has 2 LSTM layers and a softmax layer which detects nodding and yawning, the other has 3 LSTM layers and a softmax layer which detects operating steering wheel, followed by a 2-layer DNN.
- 3-3-*LSTM-DNN:* It has 2 LSTM networks with same structures, one has 3 LSTM layers and a softmax layer which detects nodding and yawning, the other detects operating steering wheel, followed by a 2-layer DNN.

The accuracy of each action under different network structures are shown in Fig. 12. It can be seen that 3-3-LSTM-DNN has the highest accuracy, but its complex structure and high computational complexity make it inappropriate to be deployed on smartphones. 2-3-LSTM-DNN and 3-LSTM-DNN also achieve high average accuracy, but the accuracy of nodding and yawning detection by 3-LSTM-DNN are comparably low. This situation may hurt the performance of D³-Guard in some cases. Thus, we choose 2-3-LSTM-DNN as our default network through empirical studies.

### F. Impact of Training Dataset Size

Since we adopt deep learning methods, the size of training dataset has great influence on system performance. Fig. 13 shows the relationship between the size of collected training dataset and the accuracy of action detections. It is obvious that with the size of training dataset increases, the accuracy of D³-Guard also increases at the beginning. But when the accuracy reaches a certain value, the accuracy does not increase with the increase of training dataset size. And according to Section IV-B, the size of training dataset is large enough to ensure the high accuracy of our system.

### G. Impact of Frame Length

Fig.14 shows the accuracy of action detections under different frame lengths. It can be seen that when the frame length is between $0.2s$ and $0.3s$, D³-Guard can achieve the highest accuracy. Fig. 15 shows the percentage that our system correctly detects a drowsy driving action within 60% of this action's total duration under different frame lengths. The percentage maintains relatively high when the frame length is shorter than $0.3s$, but decreases rapidly when the frame length is longer than $0.3s$. A small frame length which is shorter than $0.2s$ can bring high computational complexity to smartphones. Thus, we adopt 0.25-second frame length in D³-Guard.

### H. Impact of Other Factors

In our experiments, we also evaluate the impact of other factors on accuracy, such as driving while listening to the radio, driving on a crooked road and driving at different speeds ($20km/h$ and $40km/h$). Fig. 16 shows that these 4 factors have little effects on accuracy. $20kHz$ has exceeds the frequency of most sounds in nature, so noise has almost no effect on D³-Guard. Crooked road has the most obvious impact, since driver operates steering wheel more often on crooked roads, resulting in relatively low accuracy.

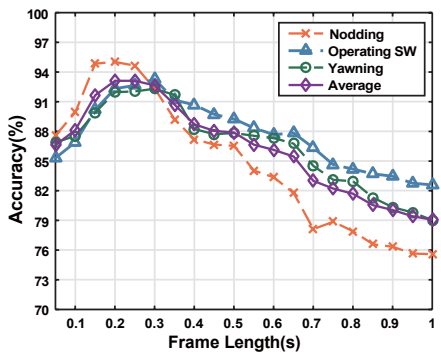
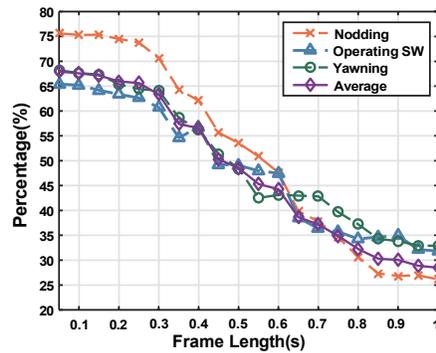
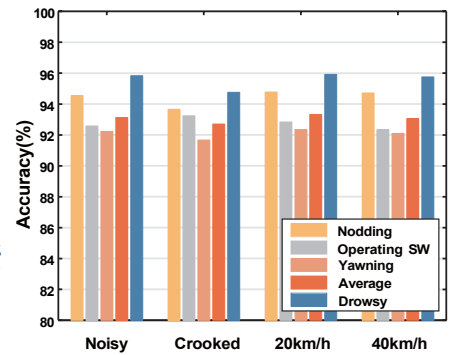

Fig. 14: Accuracy for different frame length.

Fig. 15: Timeliness for different frame length.

Fig. 16: Accuracy for other four factors.

## VI. CONCLUSION

In this paper, we address how to detect drowsy driving in real time to improve driving safety. We find that for a particular drowsy driving action, different drivers have similar patterns. We propose a real-time drowsy driving detection system, $D^3$-Guard, to detect drowsy driving behaviors at early stage which purely leverages build-in audio devices on smartphones. $D^3$-Guard extracts features of nodding, yawning and operating steering wheel based on Doppler profiles. $D^3$-Guard then takes advantages of LSTM networks to build an efficient detector which recognize the three drowsy driving actions. We conduct extensive experiments in real driving environments and the results show $D^3$-Guard works in a desirable way.

ACKNOWLEDGMENTS

The work of Fan Li is partially supported by the NSFC (No. 61772077, 61370192), and the Beijing Natural Science Foundation (No. 4192051). The work of Song Yang is partially supported by the NSFC (No. 61802018) and Beijing Institute of Technology Research Fund Program for Young Scholars. The work of Yu Wang is partially supported by the NSFC (No. 61572347), the US National Science Foundation (No. CNS-1343355), and by the U.S. Department of Transportation Center for Advanced Multimodal Mobility Solutions and Education (No. 69A3351747133). Fan Li is the corresponding author.